\documentclass[aps,prd,groupedaddress]{revtex4}
\usepackage{graphicx}
\usepackage{epsfig}
\usepackage{amsfonts}
\usepackage{amssymb}
\usepackage{epsf}
\usepackage{color}
\newcommand{\insertplot}[5]{\begin{figure}
 \hfill\hbox to 0.05in{\vbox to #5in{\vfill
 \inputplot{#1}{#4}{#5}}\hfill}
 \hfill\vspace{-.1in}
 \caption{#2}\label{#3}
 \end{figure}}
\newcommand{\inputplot}[3]{
 \special{ps: plotfile #1}
}

\newcounter{fig}

\begin{document}

\title{WORMHOLE SOLUTIONS WITH NUT CHARGE\\ IN HIGHER CURVATURE THEORIES}
\author{Rustam Ibadov}
\email[]{ibrustam@mail.ru}
\affiliation{Department of Theoretical Physics and Computer Science, Samarkand State University,
Samarkand 140104, Uzbekistan}
\author{Burkhard Kleihaus}
\email[]{b.kleihaus@uni-oldenburg.de}
\affiliation{Institute of Physics, University of Oldenburg, D-26111 Oldenburg, Germany}
\author{Jutta Kunz}
\email[]{jutta.kunz@uni-oldenburg.de}
\affiliation{Institute of Physics, University of Oldenburg, D-26111 Oldenburg, Germany}
\author{Sardor Murodov}
\email[]{mursardor@mail.ru}
\affiliation{Department of Theoretical Physics and Computer Science, Samarkand State University,
Samarkand 140104, Uzbekistan$ $}

\date{\today}
\begin{abstract}
We present wormholes with a Newman-Unti-Tamburino (NUT)
charge that arise in certain higher curvature theories,
where a scalar field is coupled to a higher curvature invariant.
For the invariants we employ i) a Gauss-Bonnet term and ii) a 
Chern-Simons term, which then act as source terms for the scalar field.
We map out the domain of existence of wormhole solutions 
by varying the coupling parameter and the scalar charge 
for a set of fixed values of the NUT charge.
The domain of existence for a given NUT charge is then
delimited by the set of scalarized nutty black holes,
a set of wormhole solutions with a degenerate throat
and a set of singular solutions.
\end{abstract}

\maketitle

\section*{Introduction}

Wormholes are intriguing solutions of numerous theories of gravity.
In General Relativity the presence of some form of exotic matter is required
in order to construct traversable Lorentzian wormholes,
since the energy conditions must be violated
(see, e.g.,~\cite{Morris:1988cz,Visser:1995cc,Lobo:2017eum}.)
The simplest possibility here is to employ a phantom scalar field,
i.e., a field whose kinetic term has the opposite sign as compared
to an ordinary scalar field.
Such phantom fields have been already employed decades ago by Ellis and
Bronnikov, when constructing wormhole solution in General Relativity
 \cite{Ellis:1973yv,Ellis:1979bh,Bronnikov:1973fh,Kodama:1978dw}.

In alternative theories of gravity, however, classical Lorentzian wormholes
may arise without the need for exotic matter.
The reason is, that the gravitational interaction itself can give rise to
additional terms, that may be interpreted as contributing to an
effective stress energy tensor on the right hand side of the 
Einstein field equations, which then may lead to violation of the energy conditions.
Thus it is the modified gravitational interaction itself
which provides the necessary violation of the energy conditions (see e.g.,
\cite{Hochberg:1990is,Fukutaka:1989zb,Ghoroku:1992tz,Furey:2004rq,Bronnikov:2009az,Kanti:2011jz,Kanti:2011yv,Lobo:2009ip,Harko:2013yb}).

In recent years alternative gravity theories have received much attention
\cite{Will:2005va,Faraoni:2010pgm,Berti:2015itd,CANTATA:2021mgk}.
To a large extent this interest has been driven by the quest to resolve
various cosmological issues. However, alternative theories of gravity
may also lead to interesting predictions for compact objects
in the strong gravity regime, where the advent of gravitational wave
multi-messenger astronomy has opened a new window into the universe
with the new observational possibilities and challenges.

Among the numerous alternative gravitational theories we consider
higher curvature theories very attractive, where a scalar field
is coupled to a higher curvature invariant.
On the one hand higher curvature terms arise under various circumstances, 
when quantum gravity theories are considered,
and on the other hand their coupling constants may still be large,
thus leading to interesting potentially observable effects
in the strong gravity regime, since constraints from the solar system
or binary pulsars may easily be satisfied.

Here we consider two such theories:
i) Einstein-scalar-Gauss-Bonnet (EsGB) theory
and ii) Einstein-scalar-Chern-Simons (EsCS) theory.
In both cases we choose the same type of coupling function
of the scalar field to the invariant,
namely a function quadratic in the scalar field.
When such a quadratic coupling is chosen, 
a vacuum solution of General Relativity supplemented with vanishing scalar field 
is also a solution of EsGB and EsCS theory.
Thus the Schwarzschild black hole and the Kerr black hole
remain solutions of these alternative gravity theories.
However, these theories may possess additional black hole solutions
with a finite scalar field, i.e., spontaneously scalarized black holes.

In the case of EsGB theory spontaneously scalarized black holes
were first obtained in the static spherically symmetric case
\cite{Antoniou:2017acq,Doneva:2017bvd,Silva:2017uqg}
and later generalized to include rotation
\cite{Cunha:2019dwb,Collodel:2019kkx,Dima:2020yac,Herdeiro:2020wei,Berti:2020kgk}.
Here the Gauss-Bonnet (GB) invariant acts as an effective mass term in the
scalar field equation, that may induce a tachyonic instability
of the Schwarzschild or Kerr black holes.

In the case of EsCS theory there are no spontaneously scalarized 
static spherically symmetric black holes.
The reason is that the Chern-Simons (CS) invariant 
vanishes for the Schwarzschild solution.
So no tachyonic instability will be induced in this case.
For the CS invariant to contribute
one would need the presence of a parity-odd source.
Such a source would be rotation \cite{Yunes:2009hc,Delsate:2018ome},
however, the construction of rotating spontaneously scalarized
EsCS black holes will be technically rather challenging.

Therefore, as a first step towards obtaining rotating spontaneously scalarized black holes
in EsCS theory, a by far simpler case has been studied in 
\cite{Brihaye:2018bgc}.
Here a NUT charge has been introduced to obtain a parity odd source term,
but retain a system of ordinary differential equations,
since the angular dependence factorizes.
The CS invariant can then act as an effective mass term in the
scalar field equation and induce a tachyonic instability
of the Schwarzschild black holes.
A NUT charge is, in fact, quite intriguing.
It gives, for instance, rise to a Misner string on the polar axis
and implies the presence of closed timelike curves
(see e.g. \cite{Kagramanova:2010bk} and references therein).

Here we consider wormholes with NUT charge in i) EsGB theory and ii) EsCS theory,
recalling and extending previous results \cite{Antoniou:2019awm,Ibadov:2020btp, Ibadov:2020ajr}.
Since one boundary of the domain of existence of wormholes
in such theories is typically given by the corresponding set of black holes,
we follow closely Brihaye et al.~\cite{Brihaye:2018bgc},
who have obtained the sets of spontaneously scalararized EsGB and EsCS 
black holes with NUT charge.
The presence of the NUT charge does not lead to asymptotically flat solutions
in the usual sense. However, the asymptotic fall-off of the functions
is of the usual type of an asymptotically flat spacetime.

The paper is organized as follows:
We first present the action for both theories and their equations of motion.
Then we discuss the conditions for the presence of a throat.
We next determine the boundary conditions, the junction conditions
and the energy conditions.
Subsequently, we present our results for i) EsGB wormholes
and ii) EsCS wormholes with NUT charge.
We then end with our conclusions.

\section*{Action and equations of motion}

We employ the effective action for Einstein-scalar-higher curvature
invariant theories as presented in Brihaye et al.~\cite{Brihaye:2018bgc}
(with geometrized units with $G=c=1$)
\begin{eqnarray}
S=\frac{1}{16 \pi}\int \left[R - \frac{1}{2}
 \partial_\mu \phi \,\partial^\mu \phi 
 + F(\phi){\cal I}(g) \right] \sqrt{-g} d^4x  \ ,
\label{act}
\end{eqnarray}
where $R$ is the curvature scalar,
$\phi$ denotes a massless scalar field without
self-interaction, coupled with coupling function
$F(\phi)$ to an invariant ${\cal I}(g)$. 
For the coupling function
$F(\phi)$ we choose a quadratic $\phi$-dependence with coupling
constant $\alpha$,
\begin{equation}
F(\phi) = \alpha \phi^2\ .
\label{FandU}
\end{equation}
This is the simplest choice that leads to spontaneous
scalarization of black holes.
We emphasize, that we follow the conventions of Brihaye et al.~\cite{Brihaye:2018bgc}, 
since we would like to compare with their results.
Obviously, any prefactor of the kinetic term in the action 
Eq.~(\ref{act}) can be absorbed by re-definitions of the 
scalar field $\phi$ and parameter $\alpha$  of the coupling function 
$F(\phi) = \alpha \phi^2$.

For the invariant ${\cal I}(g)$ we make the following choices: 
(i) we  choose the Gauss-Bonnet term
\begin{eqnarray}
{\cal I}(g)= R^2_{\rm GB} = R_{\mu\nu\rho\sigma} R^{\mu\nu\rho\sigma}
- 4 R_{\mu\nu} R^{\mu\nu} + R^2 \ ,
\end{eqnarray}
and (ii) we choose the Chern-Simons term
\begin{eqnarray}
{\cal I}(g)= R^2_{\rm CS} =
   ^\ast\!{\!\!R^\mu_{{\phantom \mu}\nu}}^{\rho\sigma} R^\nu_{{\phantom \nu}\mu\rho\sigma}
\ .
\end{eqnarray}
Here we have employed the Hodge dual of the Riemann-tensor
$^\ast{\!R^\mu_{{\phantom \mu}\nu}}^{\rho\sigma}
  = \frac{1}{2}\eta^{\rho\sigma\kappa\lambda} R^\mu_{{\phantom \mu}\nu\kappa\lambda}
$
which is defined with the 4-dimensional Levi-Civita tensor
$\eta^{\rho\sigma\kappa\lambda}=\epsilon^{\rho\gamma\sigma\tau}/\sqrt{-g}$.
We note that both invariants are topological in four dimensions.
However, their coupling to the scalar field $\phi$
lets them contribute to the equations of motion.

The coupled sets of field equations are then obtained
by variation of the action (\ref{act}) with respect to
the scalar field and the metric,
\begin{equation}
\nabla^\mu \nabla_\mu \phi  + \frac{dF(\phi)}{d\phi}{\cal I}=0 \ ,
\label{scleq}
\end{equation}
\begin{equation}
G_{\mu\nu}  = \frac{1}{2}T^{({\rm eff})}_{\mu\nu} \ .
\label{Einsteq}
\end{equation}
Here $G_{\mu\nu}$ is the Einstein tensor, as usual, while $T^{({\rm eff})}_{\mu\nu}$
denotes the resulting effective stress energy tensor
\begin{equation}
T^{({\rm eff})}_{\mu\nu} = T^{(\phi)}_{\mu\nu} +  T^{({\cal I})}_{\mu\nu} \ ,
\label{teff}
\end{equation}
consisting of the standard scalar field contribution
\begin{equation}
T^{(\phi)}_{\mu\nu} = \left(\nabla_\mu \phi\right) \left(\nabla_\nu \phi\right)
                   -\frac{1}{2}g_{\mu\nu}\left(\nabla_{\!\rho}\, \phi\right) \left(\nabla^\rho \phi\right) \ ,
\label{tphi}
\end{equation}
and a gravitational contribution from the respective higher curvature invariant ${\cal I}(g)$.
For the two above invariants we obtain (i)
%
%
\begin{equation}
T^{(GB)}_{\mu\nu} =
\left(g_{\rho\mu} g_{\lambda\nu}+g_{\lambda\mu} g_{\rho\nu}\right)
\eta^{\kappa\lambda\alpha\beta} %
\phantom{}^\ast {\!R^{\rho\gamma}_{\phantom{\rho\gamma}\alpha\beta}}
\nabla_\gamma \nabla_\kappa F(\phi) \ ,
\label{teffi}
\end{equation}
%
and (ii)
\begin{equation}
T^{(CS) \mu\nu} =
-8 \left[\nabla_\rho F(\phi)\right] \epsilon^{\rho\sigma\tau ( \mu}
                  \nabla_\tau R^{\nu )}_{\phantom{\nu )}\sigma}
                 +\left[\nabla_\rho \nabla_\sigma F(\phi)\right]
^\ast{\!\!R^{\sigma ( \mu \nu )\rho }}    .
\label{teffii}
\end{equation}

To obtain wormhole solutions with a NUT charge $N$ we assume
the line element to be of the form
\begin{equation}
ds^2 = -e^{f_0}\left(dt - 2 N \cos\theta d\varphi\right)^2 +e^{f_1}\left[dr^2
+r^2\left( d\theta^2+\sin^2\theta d\varphi^2\right) \right]\ ,
\label{met}
\end{equation}
With this ansatz the angular dependence factorizes, and 
all three functions,
the two metric functions $f_0$ and $f_1$ and the scalar field function $\phi$,
depend only on the radial coordinate $r$.

Insertion of the above ansatz (\ref{met}) for the metric and the scalar field into
the scalar field equation (\ref{scleq})
and the Einstein equations (\ref{Einsteq}) with effective
stress-energy tensor (\ref{teff})
leads to five coupled, nonlinear ordinary differential equations (ODEs).
However, these are not all independent, 
and one ODE can be treated as a constraint.
We thus retain three coupled ODEs of second order to be solved numerically.
We remark that in case (i) 
we can even obtain one first order and two second order ODEs.

The field equations possess an invariance under the scaling transformation
%
%
\begin{equation}
r \to \chi r \ , \ \ \  N \to \chi N \ , \ \ \   t \to \chi t \ , \ \ \  F \to \chi^2 F\  ,
\label{scalinvar}
\end{equation}
with constant $ \chi > 0$.

\section*{Throat and boundary conditions}

%
In order to obtain wormhole solutions, the presence of a throat is mandatory.
We therefore introduce the circumferential (or spherical) radius
\begin{equation}
R_C=e^{\frac{f_1}{2}} r \
\label{rc}
\end{equation}
of the spacetime. 
When the circumferential radius has a single finite extremum,
a minimum,
this corresponds to the single throat of the wormhole,
since the throat is a surface of minimal area.
While wormholes with more extrema also exist,
which then feature an equator surrounded by a double throat,
we here focus on single throat wormholes.

We now require the presence of an extremum
of the circumferential radius at some $r=r_0$,
in order to obtain the first set of boundary conditions.
This leads to
\begin{equation}
\left. \frac{dR_C}{dr} \right|_{r=r_0} = 0 \ \ \
\Longleftrightarrow \ \ \ \left.   \frac{df_1}{dr}\right|_{r=r_0} =-\frac{2}{r_0} \ .
\label{extr_rc}
\end{equation}
Consequently, we will refer to the two-dimensional submanifolds
defined by $r=r_0$ and $t=$const.~as the throat of the wormholes.

The presence of the NUT charge endows these wormholes with
an interesting feature:
%
the throat metric of the wormholes with NUT charge
\begin{equation}
ds^2_{\rm th} = e^{f_1(r_0)} r_0^2 \left(d\theta^2
                +\left[\sin^2\theta -\frac{4N^2}{r_0^2} e^{f_0(r_0)-f_1(r_0)} \cos^2\theta\right]d\varphi^2
        \right) \
\label{met_th}
\end{equation}
changes its signature, when the coefficient of $d\varphi^2$ changes sign.
This happens at the critical angles $\theta_c$ and $\pi-\theta_c$.
where $\theta_c$ is obtained by requiring
$\det(g_{\rm th}) \geq 0$,
and thus $\theta_c \leq \theta \leq \pi-\theta_c$
with
\begin{equation}
\theta_c= \left.
\arctan\left(\frac{2|N|}{r_0} e^{\frac{f_0-f_1}{2}}\right)
\right|_{r_0} .
\label{thetac}
\end{equation}
The metric has positive signature, as required for a two-dimensional Riemannian surface,
only for $\theta_c \leq \theta \leq \pi-\theta_c$.
The signature change is, however, not a surprise for
a spacetime with a NUT charge,
since a NUT charge leads to a non-causal structure of
a spacetime and allows for closed timelike curves.

The second set of boundary conditions is obtained by imposing 
the usual fall-off as for asymptotically flat spacetimes and thus
the usual boundary conditions for $r \to \infty$ \cite{Brihaye:2018bgc}.
Thus the metric functions and the
scalar field satisfy the asymptotic expansions
%
%
\begin{eqnarray}
f_0  & = & - \frac{2M}{r} + {\cal O} \left(r^{-3}\right) ,
\label{expf0} \\
f_1  & = &  \ \frac{2M}{r} + {\cal O} \left(r^{-2}\right) ,
\label{expf1} \\
\phi  & = & \phi_\infty - \frac{D}{r} + {\cal O} \left(r^{-3}\right) ,
\label{exphi0}
\end{eqnarray}
where the constants $M$ and $D$ denote the mass 
and the scalar charge of the wormholes, respectively,
while $\phi_\infty$ denotes the asymptotic
value of the scalar field.
All higher order terms in the expansion can be
expressed in terms of the constants $M$, $D$ and $\phi_\infty$. A solution
is therefore uniquely determined by these three constants (and,
of course, the parameters of the theory). 
Since we follow Brihaye et al.~\cite{Brihaye:2018bgc}
in order to compare with their spontaneously scalarized black hole solutions,
we need to also employ their chosen asymptotic value $\phi_\infty=0$.

\section*{Junction conditions}
Our aim is to obtain wormholes, which are symmetric with respect to the throat.
Inspection of the functions at the throat shows, that if the solutions would be
simply continued to the other side of the throat, the solutions would not be symmetric.
Worse, however, is that further integration beyond the throat will lead to a curvature
singularity, and asymptotic infinity on that side cannot be reached.
Therefore we cut the solutions at the throat
and paste a symmetric copy on the other side of the throat.
In order to have continuous and differentiable wormholes,
we therefore need to impose junction conditions at the throat.

To that end we introduce
the radial coordinate $\eta$,
\begin{equation}
\eta= r_0\left(\frac{r}{r_0}-\frac{r_0}{r}\right) \ ,
\label{defeta}
\end{equation}
where $r_0$ is a constant, and define the constant $\eta_0$ via
$\eta_0 = 2 r_0$.
We then reexpress the metric in terms of the new radial coordinate $\eta$
\begin{equation}
ds^2 = -e^{f_0}\left(dt-2N \cos\theta d\varphi\right)^2
+e^{{F}_1}\left[d\eta^2
               +\left(\eta^2+\eta_0^2\right) \left(d\theta^2+\sin^2\theta d\varphi^2\right)\right] \ ,
\label{ds2eta}
\end{equation}
and the new metric function $F_1$,
$$
e^{{F}_1} = e^{f_1}\left(1+\frac{r_0^2}{r^2}\right)^{-2} \ .
$$
Thus the throat is located at $\eta=0$, and the solution for $\eta \le 0$ is obtained from the solution for
$\eta \ge 0$ by imposing the conditions
$f_0(-\eta) = f_0(\eta)$, ${F}_1(-\eta) = {F}_1(\eta)$ and
$\phi(-\eta) = \phi(\eta)$.
Since these conditions generically introduce jumps in the derivatives of the functions $f_0$ and $\phi$
at the throat, we impose the presence of a thin shell of matter at the throat.

The proper procedure here is the make use  of
an appropriate set of junction conditions \cite{Israel:1966rt,Davis:2002gn},
namely we evaluate the jumps in the Einstein and scalar field equations
that arise for $\eta \rightarrow -\eta$,
\begin{equation}
\langle G^\mu_{\phantom{a}\nu} -T^\mu_{\phantom{a}\nu}\rangle = s^\mu_{\phantom{a}\nu} \ , \ \ \
\langle \nabla^2 \phi + \dot{F} {\cal I}\rangle = s_{\rm scal} \ .
\label{jumps}
\end{equation}
%
Here the stress energy tensor of the matter at the throat has been denoted by $s^\mu_{\phantom{a}\nu}$,
while the source term for the scalar field has been named $s_{\rm scal}$,
and $dF(\phi)/d\phi = \dot F$ denotes the derivative of the coupling function.
Since the thin shells should be composed of 
ordinary (non-exotic) matter,
we assume a perfect fluid at the throat
with pressure $p$ and energy density $\epsilon_c$. 
Moreover, we assume a scalar charge density $\rho_{\rm scal}$ together with a gravitational source
\begin{equation}
S_\Sigma = \int \left[\lambda_1 + 2 \lambda_0 F(\phi) \bar{R}\right]\sqrt{-\bar{h}} d^3 x \ ,
\label{act_th}
\end{equation}
employed before for EsGB wormholes without NUT charge \cite{Kanti:2011jz,Kanti:2011yv,Antoniou:2019awm}.
Here $\lambda_1$ and $\lambda_0$ are constants,
$\bar{h}_{ab}$ is the induced metric at the throat, and
$\bar{R}$ the corresponding Ricci scalar. 
The junction conditions are obtained by substituting the metric into the set of
equations (\ref{jumps}).

Now we evaluate the junction conditions for both invariants successively.
Note, that all functions and derivatives are evaluated at the throat.
For the Gauss-Bonnet invariant we obtain
\begin{eqnarray}
\frac{4}{\eta_0^2} \dot{F} \phi'
\left(\eta_0^2 e^{-\frac{3}{2} {F}_1} +3 N^2 e^{f_0-\frac{5}{2} {F}_1} \right)
& = &
\lambda_1\eta_0^2 + 4\lambda_0 F \frac{ \eta_0^2 e^{-{F}_1}
                               +3 N^2 e^{f_0-2{F}_1}}{\eta_0^2}
- \epsilon_c\eta_0^2 \ ,
\label{junc_gb_00}\\
N\cos\theta\left[
\eta_0^2 f_0' e^{-\frac{{F}_1}{2}}
-8\dot{F} \phi'
\left(e^{-\frac{3}{2} {F}_1}  +\frac{4 N^2}{\eta_0^2}  e^{f_0-\frac{5}{2} {F}_1}\right)
\right]
& = &
2 N \cos\theta \left[
\left(\epsilon_c + p\right) \eta_0^2
- 4\lambda_0 F \frac{ \eta_0^2 e^{-{F}_1}
                               +4 N^2 e^{f_0-2{F}_1}}{\eta_0^2}
\right] \ ,
\label{junc_gb_0p}\\
\frac{\eta_0^2 f_0'}{2} e^{-\frac{{F}_1}{2}}
                       -\frac{4 N^2}{\eta_0^2} \dot{F} \phi' e^{f_0-\frac{5}{2} {F}_1}
& = &
p \eta_0^2 +\lambda_1  \eta_0^2
- 4\lambda_0 N^2  F\frac{e^{f_0-2{F}_1}}{\eta_0^2} \ ,
\label{junc_gb_pp}\\
e^{-{F}_1}\phi' - 4 \frac{\dot{F}}{\eta_0^4} f_0'
\left(\eta_0^2 e^{-2 {F}_1}+3 N^2 e^{ f_0-3{F}_1} \right)
& = &
-4\lambda_0\frac{\dot{F}}{\eta_0^4}\left(\eta_0^2 e^{-{F}_1}
                                      + N^2 e^{f_0-2{F}_1}\right)
+\frac{\rho_{\rm scal}}{2} \ , 
\label{junc_gb_ph}
\end{eqnarray}
where the prime denotes the derivative with respect to $\eta$.
These conditions follow from the
$\left( ^t_{\phantom{a}t}\right)$,
$\left( ^t_{\phantom{a}\varphi}\right)$, and
$\left( ^\varphi_{\phantom{a}\varphi}\right)$
components of the Einstein equations and from the scalar field equation, respectively.
The $\left( ^\theta_{\phantom{a}\theta}\right)$ equation is equivalent to the
$\left( ^\varphi_{\phantom{a}\varphi}\right)$ equation, all other equations are
satisfied trivially.
The $\theta$ dependence in the
$\left( ^t_{\phantom{a}\varphi}\right)$ equation
factorizes, in fact, this equation is satisfied once
the $\left( ^t_{\phantom{a}t}\right)$ and $\left( ^\varphi_{\phantom{a}\varphi}\right)$
equations are solved.

A simple example would be pressureless matter, $p=0$.
With
\begin{equation}
\lambda_0 = \frac{\dot{F}}{F} e^{-\frac{{F}_1}{2}} \phi'\ , \ \ \
\lambda_1 =\frac{ f_0'}{2}e^{-\frac{{F}_1}{2}}\ ,
\label{lambda_ex_gb}
\end{equation}
we obtain
\begin{equation}
\epsilon_c =
\frac{f_0'}{2}e^{-\frac{{F}_1}{2}} \ .
\label{eden_ex_gb}
\end{equation}
For all of our solutions $f_0' > 0$, therefore the energy density $\epsilon_c$ is always positive for this
choice of the constants $\lambda_0$ and $\lambda_1$.

For the Chern-Simons invariant we obtain
\begin{eqnarray}
8N  \dot{F} \phi' f_0' e^{\frac{f_0}{2}}  e^{-2 {F}_1}
& = &
\lambda_1\eta_0^2 + 4\lambda_0 F \frac{ \eta_0^2 e^{-{F}_1}
                               +3 N^2 e^{f_0-2{F}_1}}{\eta_0^2}
- \epsilon_c\eta_0^2 \ ,
\label{junc_cs_00}\\
N \cos\theta f_0' \left(\eta_0^2 e^{-\frac{{F}_1}{2}}
                       -24 N \dot{F} \phi' e^{\frac{f_0}{2}-2{F}_1}\right)
& = &
2 N \cos\theta \left(\epsilon_c + p\right) \eta_0^2
- 8 N \cos\theta \lambda_0 F \frac{ \eta_0^2 e^{-{F}_1}
                               +4 N^2 e^{f_0-2{F}_1}}{\eta_0^2} \ ,
\label{junc_cs_0p}\\
\frac{f_0'}{2} \left(\eta_0^2 e^{-\frac{{F}_1}{2}}
                       -8 N \dot{F} \phi' e^{\frac{f_0}{2}-2{F}_1}\right)
& = &
p \eta_0^2 +\lambda_1  \eta_0^2
- 4\lambda_0 N^2 F \frac{e^{f_0-2{F}_1}}{\eta_0^2} \ ,
\label{junc_cs_pp}\\
e^{-{F}_1}\phi' - 4 N \frac{\dot{F}}{\eta_0^2} (f_0')^2
e^{\frac{f_0-5 {F}_1}{2}}
& = &
-4\lambda_0\frac{\dot{F}}{\eta_0^4}\left(\eta_0^2 e^{-{F}_1}
                                      +N^2 e^{f_0-2{F}_1}\right)
+\frac{\rho_{\rm scal}}{2} \ ,
\label{junc_cs_ph}
\end{eqnarray}
again from the
$\left( ^t_{\phantom{a}t}\right)$,
$\left( ^t_{\phantom{a}\varphi}\right)$,
$\left( ^\varphi_{\phantom{a}\varphi}\right)$
components of the Einstein equations and the scalar field equation, respectively.

Considering again pressureless matter, $p=0$,
with
\begin{equation}
\lambda_0 = \frac{\eta_0^2 \dot{F}}{N F} e^{-\frac{f_0}{2}} \ , \ \ \
\lambda_1 =-\frac{ f_0'}{2}e^{-\frac{{F}_1}{2}}\ ,
\label{lambda_ex_cs}
\end{equation}
we now obtain
\begin{equation}
\epsilon_c =
\frac{f_0'}{2 N}\left(N e^{-\frac{{F}_1}{2}}
                     + 8\dot{F}\phi'e^{-\frac{f_0}{2}-{F}_1}\right)
       +\frac{4 N}{\eta_0^2}\dot{F}\phi'f_0' e^{\frac{f_0}{2}-2{F}_1}         \ .
\label{eden_ex_cs}
\end{equation}

\section*{Energy conditions}

In order to obtain wormhole solutions the null energy condition (NEC)
\begin{equation}
T_{\mu\nu} n^\mu n^\nu \geq 0 \
\label{NEC}
\end{equation}
must be violated. Here
$n^\mu$ may be any null vector ($n^\mu n_\mu=0$).
Consequently, all one has to show is that null vectors exist, such that
$T_{\mu\nu} n^\mu n^\nu < 0$ somewhere in the spacetime.
Choosing the null vector $n^\mu$ 
\begin{equation}
n^\mu=\left(1,\sqrt{-g_{tt}/g_{\eta\eta}},0,0\right) \ ,
\end{equation}
and thus $n_\mu=\left(g_{tt},\sqrt{-g_{tt}\,g_{\eta\eta}},0,0\right)$,
the NEC leads to
\begin{equation}
T_{\mu\nu}n^\mu n^\nu=T^t_t n^t n_t + T^\eta_\eta n^\eta n_\eta
=-g_{tt}\,(-T^t_t +T^\eta_\eta)   \ .
\end{equation}
This shows that the NEC is violated when
\begin{equation}
 -T_t^t + T_\eta^\eta  <  0 \ .
\label{nec1}
\end{equation}
Alternatively, choosing the null vector
\begin{equation}
n^\mu=\left(1,0,\sqrt{-g_{tt}/g_{\theta \theta}},0\right) \ ,
\end{equation}
leads to NEC violation when
\begin{equation}
-T_t^t + T_\theta^\theta < 0 \ .
\label{nec2}
\end{equation}
These conditions have been investigated before for various
wormhole solutions
(see e.g., \cite{Morris:1988cz,Kanti:2011jz,Kanti:2011yv,Antoniou:2019awm}).

\section*{Results}

We solve the coupled Einstein and scalar field equations numerically.
To this end we introduce the inverse radial coordinate $x=1/r$.
Consequently, the asymptotic region $r\to \infty$ corresponds to the region $x\to 0$.
Here the expansion of the metric functions and the
scalar field then reads  (Eqs.~(\ref{expf0})-(\ref{exphi0}))
%
\begin{equation}
f_0  = - 2 M x + {\cal O}\left(x^3\right) , \ \ \
f_1   =   2 M x + {\cal O}\left(x^2\right) , \ \ \
\phi   =  \phi_\infty - D x + {\cal O}\left(x^3\right) .
\label{exphi}
\end{equation}
To solve the system of ODEs, we treat it as an initial value problem,
and employ the fourth order Runge Kutta method.
The above expansion then yields the initial values,
%
%
\begin{equation}
f_{0,{\rm ini}}= 0\ , \ \ \ f'_{0,{\rm ini}}=- 2M\ , \ \ \
f_{1,{\rm ini}}= 0\ , \ \ \ f'_{1,{\rm ini}}= 2M\ , \ \ \
\phi_{{\rm ini}}= 0\ , \ \ \ \phi'_{{\rm ini}}= -D \ , 
\label{inicond}
\end{equation}
where the prime denotes the derivative with respect to $x$.
We start the computation at spatial infinity, $x=0$, and end at
the throat at some finite value $x=x_0$, where we reach the condition
(\ref{extr_rc}).

By following the outlined numerical procedure we have constructed numerous sets of
wormhole solutions with NUT charge for both invariants ${\cal I}(g)$.
In the following we demonstrate our results for both invariants successively.

\subsection*{Case i): EsGB wormholes}

\begin{figure}[t!]
\begin{center}
(a) \includegraphics[width=.45\textwidth,angle=0]{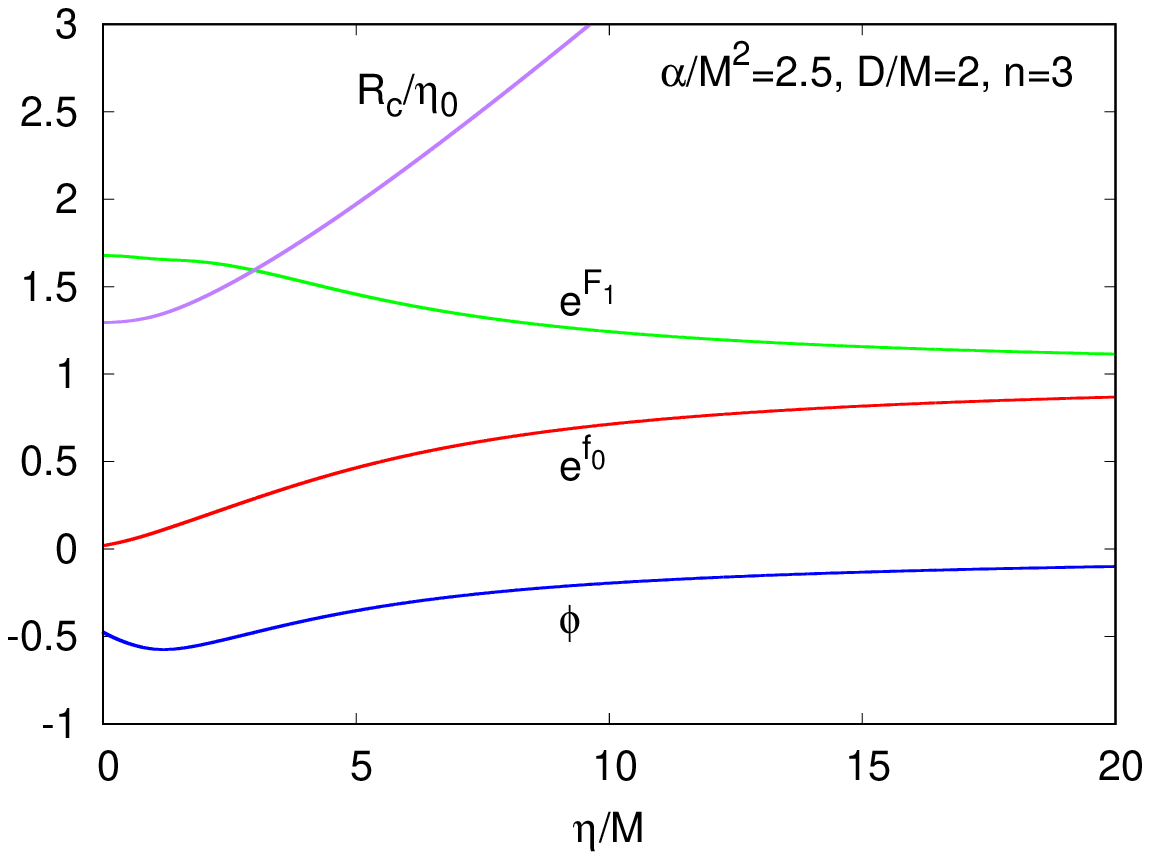} 
(b) \includegraphics[width=.45\textwidth,angle =0]{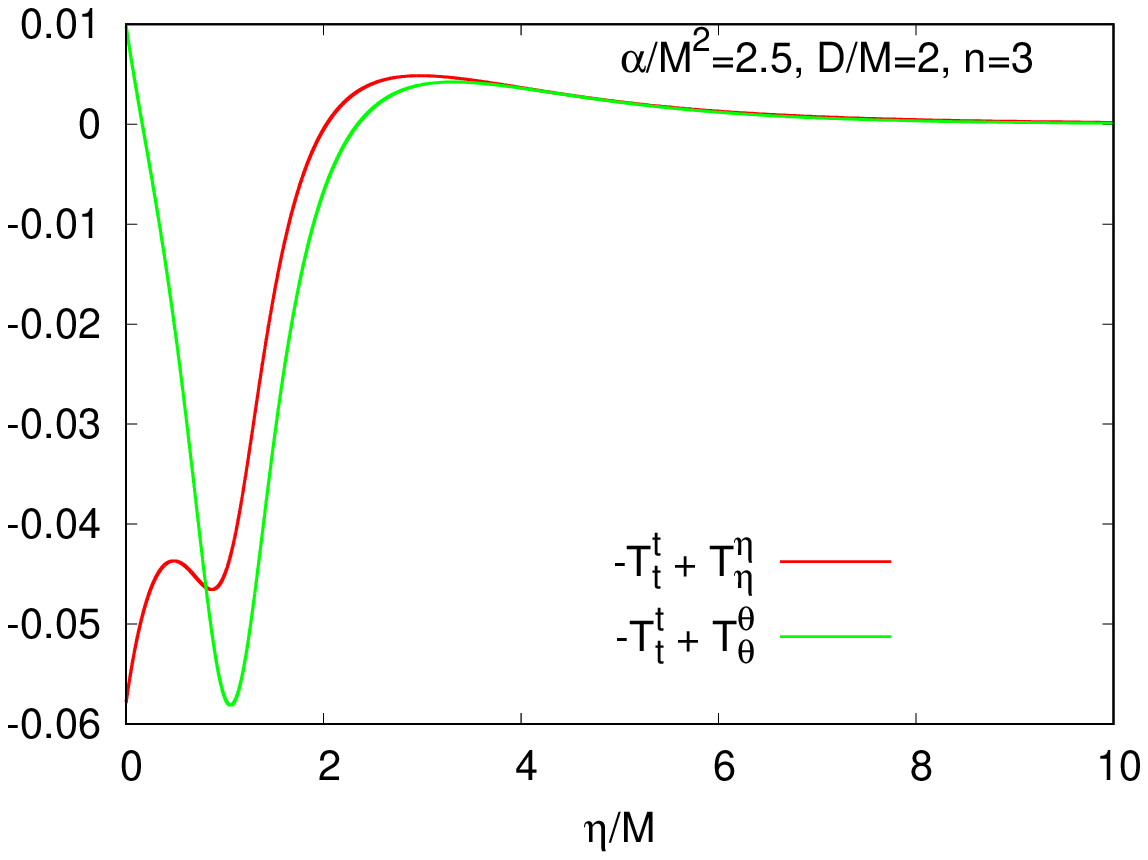}\\
(c) \includegraphics[width=.45\textwidth, angle =0]{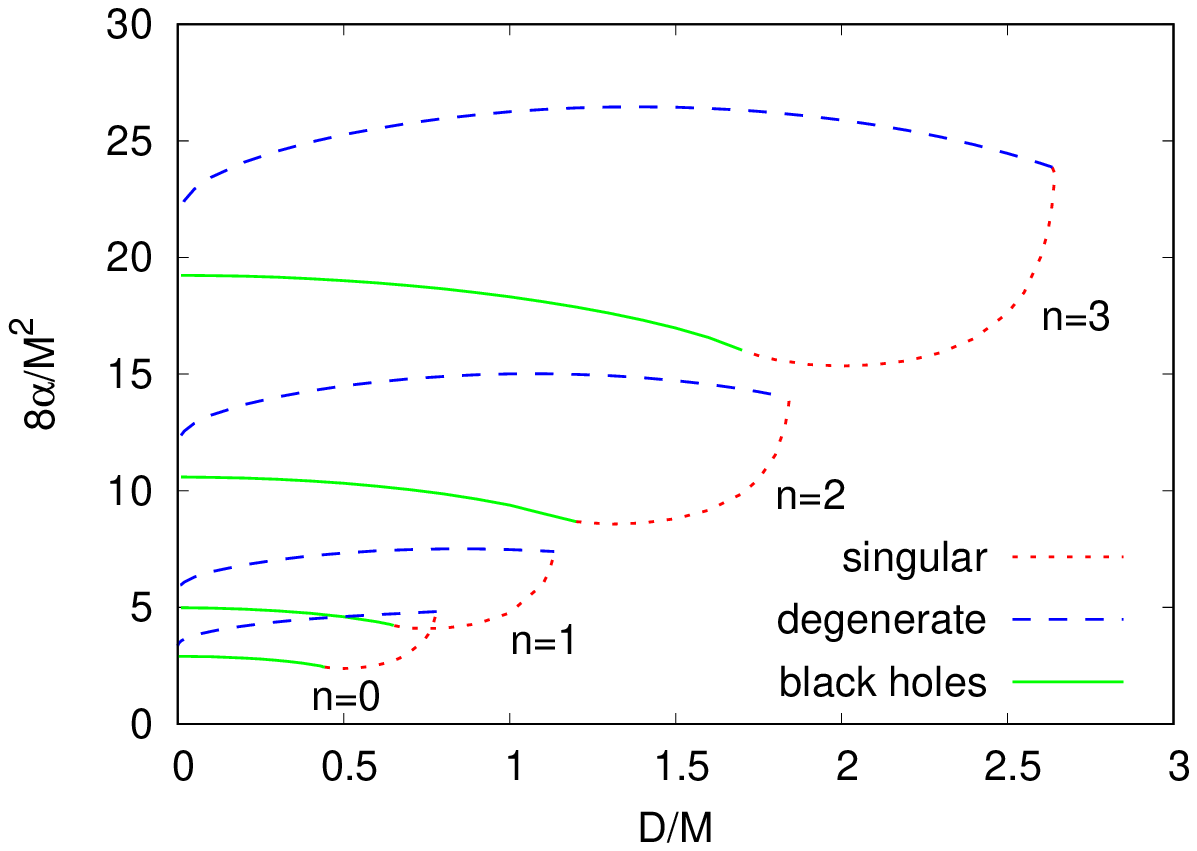}
(d) \includegraphics[width=.45\textwidth, angle =0]{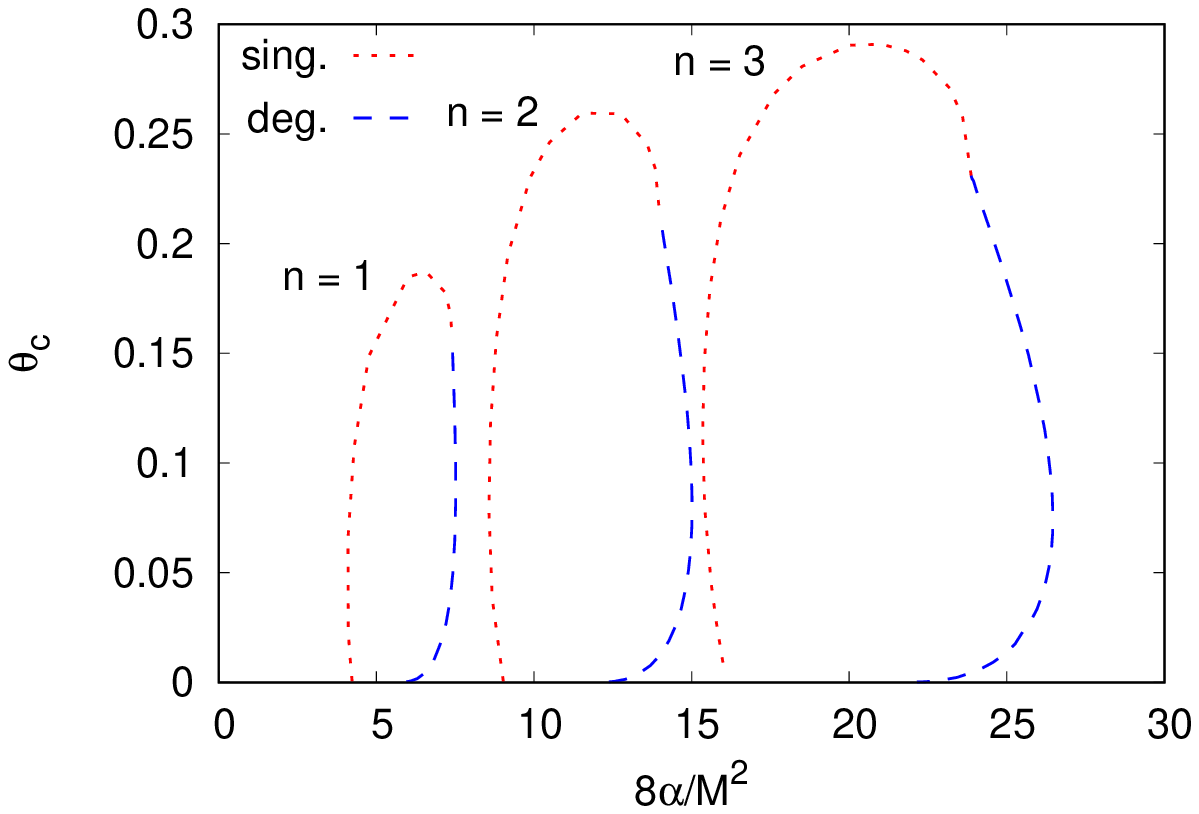} 
\end{center}
\caption{ EsGB wormhole solutions with NUT charge: 
(a) metric profile functions $e^{f_0}$, $e^{F_1}$, scalar field function $\phi$, and
scaled circumferential radius $R_c/\eta_0$ vs radial coordinate
(parameters $\alpha/M^2=2.5$, $D/M=2$ and $n=N/M=3$);
(b) NEC conditions $-T_t^t + T_\eta^\eta \geq 0$ and $-T_t^t +
T_\theta^\theta \geq 0$ vs radial coordinate $\eta$
(parameters $\alpha/M^2=2.5$, $D/M=2$ and $n=N/M=3$);
(c) domain of existence for several values of the scaled NUT charge $n=N/M$:
scaled coupling constant $8\alpha/M^2$ vs
scaled scalar charge $D/M$
(solid green: black hole limit, 
dashed blue: degenerate wormhole limit,
dotted red: singular limit);
(d) critical angle $\theta_c$ vs
scaled coupling constant $8\alpha/M^2$ 
(dashed blue: degenerate wormhole limit,
dotted red: singular limit).
}
\label{fig_solGB}
\end{figure}

We begin by showing a typical solution for EsGB wormholes with NUT charge,
where we have chosen the scaled coupling constant $\alpha/M^2=2.5$, 
the scaled scalar charge $D/M=2$ and the scaled NUT charge $n=N/M=3$.
The metric functions $e^{f_0}$ and  $e^{F_1}$, and the
scalar field function $\phi$ are shown in Fig.~\ref{fig_solGB}(a) 
versus the scaled radial coordinate $\eta/M$.
Also shown is the scaled circumferential radius $R_c/\eta_0$
versus $\eta/M$, where $\eta_0/M=0.517$. 
The minimum of $R_c$ at the coordinate origin $\eta=0$
corresponds to the location of the throat. 

As required the wormhole solutions violate the energy conditions.
We demonstrate the violation of the NEC conditions
$ -T_t^t + T_\eta^\eta \geq 0$
and
$-T_t^t + T_\theta^\theta \geq 0$
for the same solution in Fig.~\ref{fig_solGB}(b).
Clearly, both expressions are negative in a region close to the throat.
Note, that we have omitted in the notation of the energy conditions, 
that we refer to the effective stress energy tensor.
We emphasize again that the violation is caused by the
gravitational part of the effective stress energy tensor.

We next turn to the domain of existence of these EsGB wormholes
with NUT charge.
The domain of existence is shown in terms of the scaled coupling constant $8 \alpha/ M^2$
and the scaled scalar charge $D/M$ in Fig.~\ref{fig_solGB}(c)
for several values of the scaled NUT charge, $n=N/M=1$, 2, and 3,
and compared with the case of vanishing NUT charge, $n=0$ 
\cite{Antoniou:2019awm}.
Note, that we only show the right hand side $D\ge 0$ of the domain,
since the domain is symmetric with respect to $D \to -D$,
which is a consequence of the quadratic coupling function and the
chosen boundary condition $\phi_\infty=0$.

For any given value of the NUT charge
the domain of existence is delimited by three sets of solutions.
The first set corresponds to the scalarized black holes with NUT charge 
of Brihaye et at.~\cite{Brihaye:2018bgc},
which we fully reproduce. These black holes are shown by a solid green curve.
The second set is shown by a dotted red curve
and corresponds to a set of singular solutions.
Here a cusp singularity is encountered at some value 
$\eta_\star$ of the radial coordinate.
Cusp singularities are a recurring feature of scalarized wormholes
(see \cite{Antoniou:2019awm,Kleihaus:2019rbg,Kleihaus:2020qwo,Ibadov:2020btp}).
They form, when the determinant of the coefficients of the second order terms
of the ODEs vanishes.

The third set of boundary solutions is shown by a dashed blue curve
and referred to as degenerate set of solutions.
To clarify the physical meaning of this set, we recall that we identified
the throat by looking for a minimum of the circumferential radius.
However, it may happen that this extremum becomes a degenerate
extremum. Here, in principle, a new set of wormholes emerges, which possess
an equator and a double throat, i.e., the degenerate extremum
will split into a maximum (equator) and two minima (throats).
We remark, that for $D=0$ the scalar field vanishes, 
thus vacuum solutions of General Relativity are recovered.
Overall we note, that with increasing NUT charge
the domain of existence of wormhole solutions increases rapidly.

We now turn to the causal structure of the EsGB wormhole spacetimes.
The relevant quantity to consider here is the polar angle,
where the $g_{\varphi\varphi}$ component changes sign,
and thus closed timelike curves will be allowed.
In particular, we are interested in the  critical polar angle $\theta_c$,
where this change of sign occurs at the wormhole throat.
We show the boundary of the domain with changed causal structure
at the throat
in Fig.~\ref{fig_solGB}(d), where we exhibit the
critical polar angle $\theta_c$ versus 
the scaled coupling constant $8 \alpha/ M^2$
for several values of the scaled NUT charge, $n=N/M=1$, 2, and 3.
For a fixed value of the NUT charge
the dotted blue curve shows the singular boundary set,
while the dashed green curve shows the degenerate boundary set.
Only for the limiting case of black holes the critical polar angle $\theta_c$
goes to zero. Thus all the wormholes with NUT charge
possess a region with closed timelike curves
that extends to their throat.

\subsection*{Case ii): EsCS wormholes}

\begin{figure}[h!]
\begin{center}
(a) \includegraphics[width=.45\textwidth,angle=0]{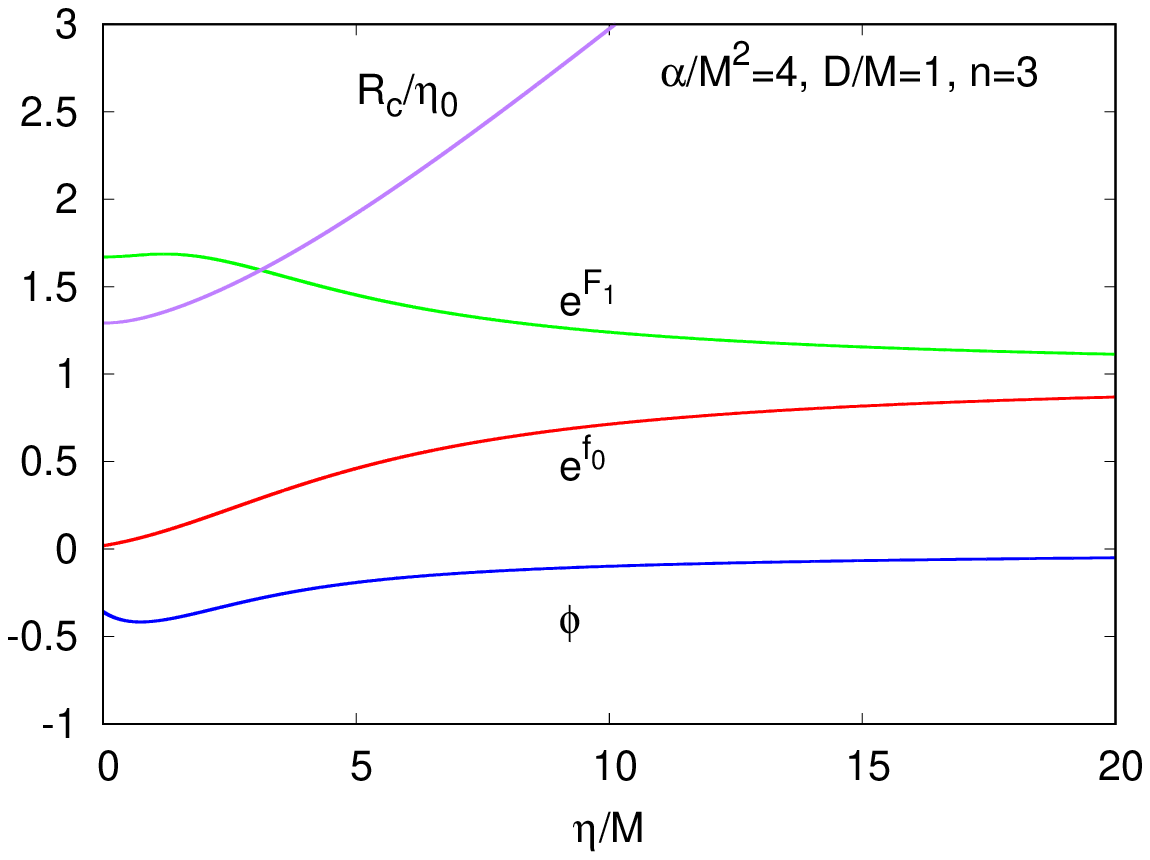} 
(b) \includegraphics[width=.45\textwidth,angle =0]{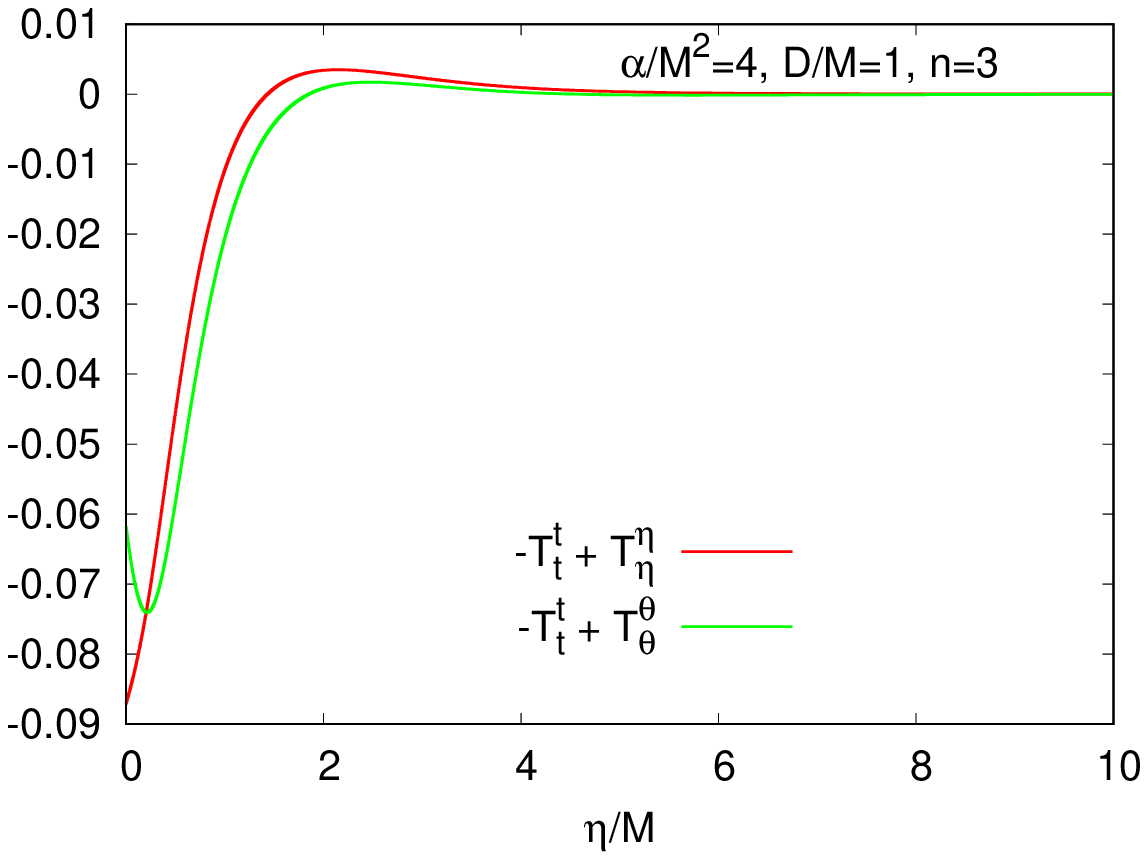}\\
(c) \includegraphics[width=.45\textwidth, angle =0]{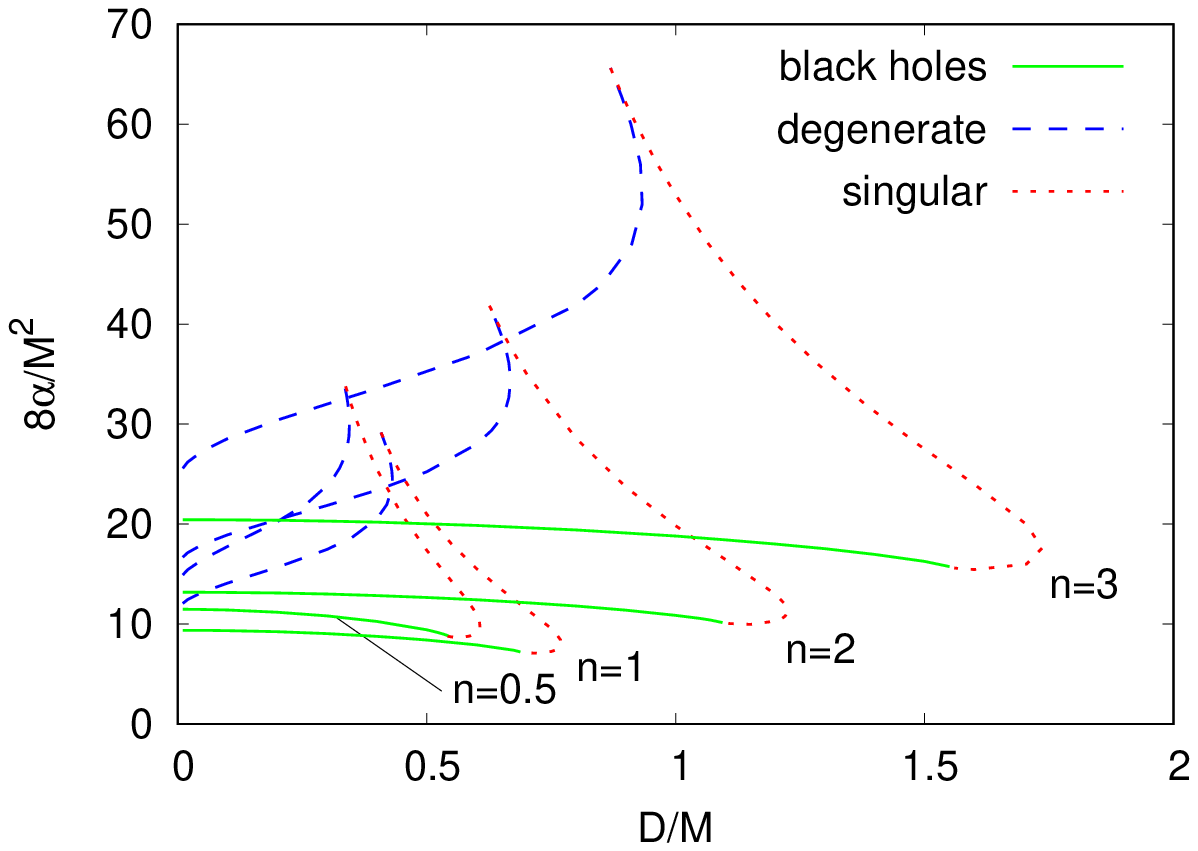}
(d) \includegraphics[width=.45\textwidth, angle =0]{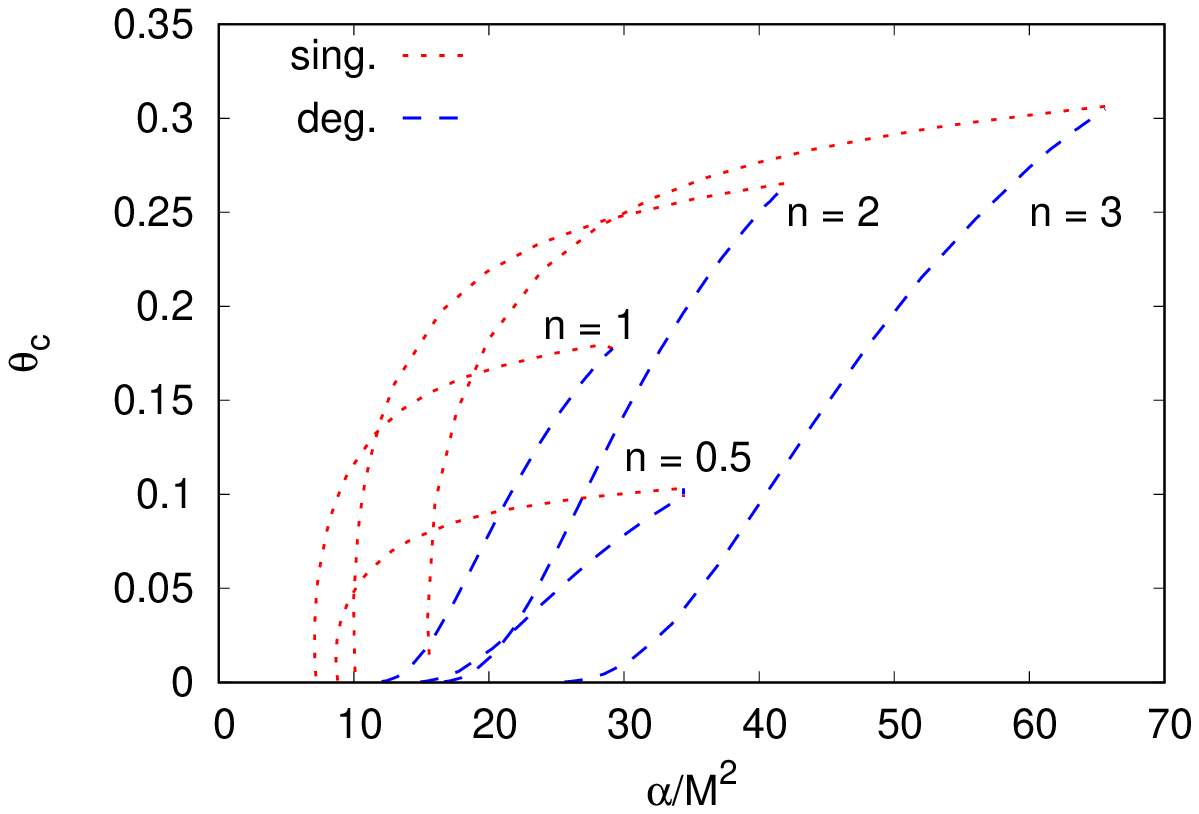} 
\end{center}
\caption{ EsCS wormhole solutions with NUT charge: 
(a) metric profile functions $e^{f_0}$, $e^{F_1}$, scalar field function $\phi$, and
scaled circumferential radius $R_c/\eta_0$ vs radial coordinate
(parameters $\alpha/M^2=4$, $D/M=1$ and $n=N/M=3$);
(b) NEC conditions $-T_t^t + T_\eta^\eta \geq 0$ and $-T_t^t +
T_\theta^\theta \geq 0$ vs radial coordinate $\eta$
(parameters $\alpha/M^2=4$, $D/M=1$ and $n=N/M=3$);
(c) domain of existence for several values of the scaled NUT charge $n=N/M$:
scaled coupling constant $8\alpha/M^2$ vs
scaled scalar charge $D/M$
(solid green: black hole limit, 
dashed blue: degenerate wormhole limit,
dotted red: singular limit);
(d) critical angle $\theta_c$ vs
scaled coupling constant $8\alpha/M^2$ 
(dashed blue: degenerate wormhole limit,
dotted red: singular limit).
}
\label{fig_solCS}
\end{figure}

We now discuss the properties of EsCS wormholes with NUT charge.
We start again with a typical wormhole solution, that has been obtained
for the parameter choice $\alpha/M^2=4$, $D/M=1$ and $n=N/M=3$.
We show its metric functions $e^{f_0}$ and $e^{F_1}$, and its
scalar field function $\phi$ versus the scaled radial coordinate $\eta/M$
in Fig.~\ref{fig_solCS}(a),
along with its scaled circumferential radius $R_c/\eta_0$,
where $\eta_0/M=0.495$. We note, that the functions are rather
similar to those of the EsGB wormholes shown in Fig.~\ref{fig_solGB}(a).

We demonstrate the violation of the NEC conditions
$ -T_t^t + T_\eta^\eta \geq 0$
and
$-T_t^t + T_\theta^\theta \geq 0$
in Fig.~\ref{fig_solCS}(b) for this solution.
Again we find a region close to the throat,
where both conditions are violated.

The domain of existence of the EsCS wormholes is shown 
in terms of the scaled coupling constant $8 \alpha/ M^2$
and the scaled scalar charge $D/M$
in Fig.~\ref{fig_solCS}(c)
for several values of the scaled NUT charge,
$n=N/M=0.5$, 1, 2, and 3.
The boundaries of the domain of existence
are completely analogous to the previously discussed case
of EsGB wormholes with NUT charge.

The first set of boundary solutions corresponds 
to the scalarized black holes with NUT charge 
of Brihaye et at.~\cite{Brihaye:2018bgc}
and is shown by a solid green curve.
The second set of boundary solutions corresponds to the
set of solutions with a cusp singularity
and is shown by a dotted red curve.
Finally the third set of boundary solutions 
corresponds to wormhole solutions
with a degenerate throat,
beyond which wormholes with an equator and a
double throat arise. They are shown 
by a dashed blue curve.
The domain of existence is again symmetric
with respect to $D \to -D$, and for $D=0$
solutions of General Relativity are recovered.

However, we also notice a new feature of the EsCS wormholes
as compared to the EsGB wormholes.
Whereas for EsGB wormholes the domain of existence
moves continuously to smaller values of the coupling constant $\alpha$,
as the NUT charge is decreased towards zero,
this is not the case for the EsCS wormholes.
Here the domain of existence exhibits
the same dependence as the scalarized EsCS black holes,
namely after reaching a minimum of $\alpha$ with decreasing
NUT charge, a further decease of the NUT charge
leads to increasing values of $\alpha$.
This is seen in Fig.~\ref{fig_solCS}(c),
where the domain of existence for $n=0.5$
has moved to larger values of $\alpha$ as compared
to the domain for $n=1$.
As the NUT charge tends towards zero,
the coupling constant $\alpha$ tends towards infinity.
However, in the limit a finite source term for the scalar field results,
since the NUT charge and the coupling constant conspire appropriately.

We finally turn to the causal structure of the EsCS wormhole spacetimes.
We show the critical polar angle $\theta_c$,
where the metric component $g_{\varphi\varphi}$ changes sign at the wormhole throat, 
versus the scaled coupling constant $8 \alpha/ M^2$
in Fig.~\ref{fig_solCS}(d)
for the same set of values of the scaled NUT charge,
$n=N/M=0.5$, 1, 2, and 3.
Again, for a fixed NUT charge, the boundaries are given by 
the dotted blue curve showing the singular boundary set,
and the dashed green curve showing the degenerate boundary set.
Only for the limiting case of black holes the critical polar angle $\theta_c$
goes to zero.

\section*{Conclusions}

We have considered scalarized wormholes with NUT charge
in two alternative gravity theories, EsGB theory and EsCS theory.
In both cases a scalar field has been coupled with a quadratic
coupling function to the respective invariant,
since this choice of coupling function allows for spontaneously
scalarized black holes with NUT charge,
i.e., solutions obtained before by Brihaye et al.~\cite{Brihaye:2018bgc}.

In these alternative gravity theories wormholes arise
without the need of introducing exotic matter.
The reason is the presence of gravitational terms 
in the stress energy tensor that result, respectively,
from the higher curvature GB term or CS term.
These terms allow for violations of the energy conditions
and therefore the presence of wormholes.

We have mapped out the domain of existence of these wormholes
for various values of the NUT charge.
In all cases, for a given NUT charge, the domain has 
one boundary consisting of scalarized black holes with NUT charge,
one boundary formed by solutions with a cusp singularity,
and one boundary composed of wormholes with a degenerate throat,
where a new type of wormhole arises,
that features an equator and a double throat,
i.e., wormholes to be addressed in the future.

In order to have symmetric wormholes without singularities
(apart from the Misner string caused by the NUT charge),
we have cut the wormhole spacetimes at the throat,
retaining only the part extending to infinity,
and pasted the symmetrically reflected copy at the throat. 
We have shown that the resulting junction conditions at the throat
can be fulfilled by a thin shell of ordinary matter such as dust, for example.

Since these wormholes carry NUT charge, 
they possess closed timelike curves.
Therefore we have investigated, in particular, the causal structure at the throat.
Interestingly, 
all these wormholes with NUT charge possess closed timelike
curves that extend all the way to the throat.
However, for the limiting set of scalarized black holes with NUT charge
the closed timelike curves do not extend to the horizon,
but can get arbitrarily close to it.

We have considered wormholes with a NUT charge
as a toy model for rotating wormholes in EsGB theory and EsCS theory.
In this toy model the angular dependence factorizes,
leaving a set of nonlinear coupled ODEs to be solved numerically.
In the case of rotating wormholes, we will have to solve sets of
nonlinear coupled partial differential equations, which will be
rather challenging for these alternative gravity theories.
Moreover, the proper conditions for the throat 
and the proper set of junction conditions
will have to be formulated and solved in this case, as well.

\section*{Acknowledgement}

BK and JK gratefully acknowledge support by the
DFG Research Training Group 1620 {\sl Models of Gravity}
and the COST Actions CA15117 and CA16104.

\end{document}